\documentclass[aps,nofootinbib,prd,eqsecnum,twocolumn,showpacs,showkeys,preprintnumbers]{revtex4-1}
\usepackage{graphicx}
\usepackage{amsmath}
\usepackage{amsfonts}
\usepackage{amssymb}
\usepackage{color}
\usepackage{bm}
\usepackage{float}
\usepackage{mathrsfs}
\usepackage{epstopdf}
\usepackage{url}
\usepackage{footnote}
\usepackage{textcomp}
\usepackage{placeins}
\usepackage{soul}
\usepackage[normalem]{ulem}
\usepackage{esint}
\usepackage[unicode=true, pdfusetitle,
 bookmarks=true,bookmarksnumbered=false,bookmarksopen=false,
 breaklinks=false,pdfborder={0 0 1},backref=false,colorlinks=false]{hyperref}
\usepackage{multirow}
\usepackage{pifont}
\usepackage{times}
\usepackage[english]{babel}

\usepackage{float}
\usepackage{enumerate}
\usepackage{hyperref}
\usepackage{tabularx}
\makeatletter
\newcommand{\stkout}[1]{\ifmmode\text{\sout{\ensuremath{#1}}}\else\sout{#1}\fi}

\newcolumntype{L}[1]{>{\hsize=#1\hsize\raggedright\arraybackslash}X}%
\newcolumntype{R}[1]{>{\hsize=#1\hsize\raggedleft\arraybackslash}X}%
\newcolumntype{C}[1]{>{\hsize=#1\hsize\centering\arraybackslash}X}%
\newcommand*\patchAmsMathEnvironmentForLineno[1]{%
 \expandafter\let\csname old#1\expandafter\endcsname\csname #1\endcsname
 \expandafter\let\csname oldend#1\expandafter\endcsname\csname end#1\endcsname
 \renewenvironment{#1}%
   {\linenomath\csname old#1\endcsname}%
   {\csname oldend#1\endcsname\endlinenomath}}%
\newcommand*\patchBothAmsMathEnvironmentsForLineno[1]{%
 \patchAmsMathEnvironmentForLineno{#1}%
 \patchAmsMathEnvironmentForLineno{#1*}}%
\AtBeginDocument{%
\patchBothAmsMathEnvironmentsForLineno{align}%
\patchBothAmsMathEnvironmentsForLineno{flalign}%
\patchBothAmsMathEnvironmentsForLineno{alignat}%
\patchBothAmsMathEnvironmentsForLineno{gather}%
\patchBothAmsMathEnvironmentsForLineno{multline}%
}
\begin{document}
\title{Study of the cosmological tensions and DESI-DR2 in the framework of the Little Rip model}
\author{Safae Dahmani$^{1,2}$}
\email{dahmani.safae.1026@gmail.com}
\author{Imad El Bojaddaini$^{1,2}$}
\email{i.elbojaddaini@ump.ac.ma}
\author{Amine Bouali$^{1,2,3}$}
\email{a1.bouali@ump.ac.ma}
\author{Ahmed Errahmani$^{1,2}$}
\email{ahmederrahmani1@yahoo.fr}
\author{Taoufik Ouali$^{1,2}$}
\email{t.ouali@ump.ac.ma}

\date{\today }
\affiliation{
$^{1}$Laboratory of Physics of Matter and Radiation, University of Mohammed first, BP 717, Oujda, Morocco\\
$^{2}$Astrophysical and Cosmological Center, Faculty of Sciences, University of Mohammed first, BP 717, Oujda, Morocco\\
$^{3}$Higher School of Education and Training, Mohammed I University, BP 717, Oujda, Morocco}
\begin{abstract}
We present an analysis that investigates the $H_0$ and $S_8$ tensions by considering a dark energy model. The latter is a late-time model characterized by a future abrupt event known as the Little Rip (LR) model and characterised by one extra parameter, $\beta$, compared to the standard model, $\Lambda$CDM. To test this approach, we perform a statistical analysis by the MCMC method using the most recent observational data. We obtain a positive correlation in ($H_0$, $\beta$) plane.   We also note that the Hubble tension is less than $3\sigma$ when using early measurements, i.e., Cosmic Microwave Background (CMB) data, and when combining it with Baryon Acoustic Oscillation (BAO) data, but it is no longer so when we combine early and late measurements (i.e. PantheonPlus (PP)).    In addition, we test the model with DESI-DR2 combined with CMB and recent SNIa measurements. We notice that our model shifts toward the quintessence field. For a complete statistical analysis, we use the Akaike Information Criteria and Bayesian analysis of the evidence.  According to Bayes factors, we find that the LR model provides an improved fit only to CMB data.
\end{abstract} 
\maketitle
\section{Introduction}

The study of the expansion of the Universe has been one of the most fascinating and challenging areas of research in modern cosmology. Over the past few decades, numerous observations and experiments have provided strong evidence for the accelerated expansion of the universe \citep{Riess_1998,Perlmutter_1999}. This discovery not only revolutionized our understanding of the cosmos but also raised intriguing questions about the nature of the underlying physical processes driving this expansion.\\
To explain the accelerated expansion, the concept of dark energy was introduced, representing an unknown form of energy that permeates the universe. Dark energy (DE) is postulated to possess negative pressure, which counteracts the gravitational attraction of matter, causing the expansion of the Universe to accelerate. The prevailing model used to describe this accelerated expansion is the Lambda Cold Dark Matter ($\Lambda$CDM) model, in which dark energy is attributed to the cosmological constant $\Lambda$. It is currently the most compatible model with experimental data. It assumes that the Universe has zero spatial curvature, and a nonzero cosmological constant responsible of the expansion of the Universe, and is filled today with cold dark matter (CDM), baryons and small fractions of radiation.\\
Despite the fact that the $\Lambda$CDM model has demonstrated to offer an excellent match to a wide range of cosmological data \citep{Riess_1998,Perlmutter_1999,Dunkley_2011,Hinshaw_2013,Ade_2014,Story_2015,2016_P,Alam_2017,
Troxel_2018,aghanim2020planck}, there are considerable tensions between the values of cosmological parameters derived from various datasets. The difference between the Hubble constant as determined by Cosmic Microwave Background (CMB) data and cosmic distance ladder measurements is the most glaring. The former has been determined by the Planck collaboration that reported $H_0=67.37\pm 0.54$kms$^{-1}$Mpc$^{-1}$ \citep{aghanim2020planck}, which is at $\sim 5 \sigma$ difference to the value determined by the Hubble Space Telescope (HST), $H_0=73.04\pm 1.04$kms$^{-1}$Mpc$^{-1}$ \citep{Riess_2022}.\\
The hypothesis that the $H_0$ tension could be the first indication of physics beyond $\Lambda$CDM is receiving more and more attention currently. While the current value of pressure (p) is close to negative energy density ($\rho$), it is possible to address the cosmic tension problem.  This solution can be achieved by postulating an equation of state (EoS) in the form of $p = w \rho$, with $w$ being close to $-1$.  Recently, new data from Dark Energy Spectroscopic Instrument (DESI) \citep{DESII} have generated significant interest in the scientific community, favoring dynamical dark energy over the well-known cosmological constant. This encourages the study of a wide range of dynamical dark energy models \citep{DDE1,DDE2,DDE3,DDE4,DDE5,DDE6,DDE7,DDE8,DDE9,DDE10,CP1,CP2,CP3,CP4,CP5,CP6,CP7,CP8,CP9}. \\
Similarly, the $S_8$ tension refers to the disagreement between the measurements of the amplitude of matter fluctuations on large scales obtained from CMB data and weak gravitational lensing surveys.  In fact, the $S_8$ parameter obtained from the Planck18 data is constrained to be $S_8 = 0.832 \pm 0.013$ \citep{aghanim2020planck}. However, local measurements reveal smaller values, such as $S_8 = 0.762^{+0.025}_{-0.024}$, as observed in the combined analysis of KV450 (KiDS+VIKING-450) and DES-Y1 (Dark Energy Survey Year 1) combined \citep{Joudaki_2020}.\\
Several theoretical proposals have been studied in this framework to resolve or at least attenuate this tension. We mention DM-DE interaction \citep{Lucca_2020,Kumar_2016,Valentino_2017}, massive neutrinos \citep{Dahmani_2023_1} and Decaying Dark Matter \citep{Kanhaiya_2020}. In addition, it has been shown that a reconstruction of dark energy  may help alleviate this ailment like early dark energy and dynamical dark energy \citep{E1, E2, E3, E4, E5} (see also \citep{E7, E8, E9,George_2020,Camarena_2021,Benevento_2020}.\\
One class of models that have been used to study these tensions is known as phantom models, which deviate from the standard $\Lambda$CDM model by allowing for exotic forms of dark energy that violate the null energy condition. These models provide an alternative viewpoint on the universe's expansion and offer a pathway to explore potential solutions for the $H_0$ and $S_8$ tensions. The authors of this paper have tested the so-called the Little Sibling of the Big Rip model (LSBR) in a recent published study \citep{Dahmani_2023_2}. 
Phantom dark energy models, which appear to be supported by observation \citep{E1, Vagnozzi_2018}, could be a good alternative for $\Lambda$CDM. \\
The Little Rip (LR) model considered in the present paper has gained particular attention due to its intriguing properties and motivations. The LR model extends the concept of the cosmic doomsday, where the Universe continues to expand indefinitely, resulting in a continuous acceleration without a future singularity. This model features a dark energy component characterized by an equation of state parameter $w$ that decreases with time, eventually approaching the phantom divide, where $w$ reaches -1. In the Little Rip scenario, the Universe gradually becomes dominated by dark energy, leading to a "rip" in the fabric of spacetime \citep{Stefancic_2005,Nojiri_d71,Nojiri_d72,Frampton_2011}. The LR scenario is a special case of the model proposed by Barrow \cite {LLL} to expand our understanding of the different types of inflationary behaviour, assuming the following general equation of state $P(\rho)=-\rho-\beta\rho^{\lambda}$, where $\beta$ and $\lambda$ are constants\footnote{A recent study shows that observational data estimates the value of $\lambda$ to $\sim-18$ \citep{LL1}.}. As a solution to the problem of the late time apocalypse without a future singularity, the LR model is defined by the following EoS, $P(\rho)=-\rho-\beta\rho^{1/2}$, by setting  $\lambda$ to $1/2$ \citep{LLLL, Stefancic_2005,Nojiri_d71,Nojiri_d72}.\\
The LR model holds significant importance within the field of cosmology. Firstly, it offers a potential resolution to the Cosmic Coincidence Problem by allowing the dark energy density to dominate as the universe approaches its "rip," removing the need for fine-tuning between dark energy and matter densities. Secondly, it provides an alternative fate for the Universe compared to the Big Rip scenario, enabling the possibility of a continuous acceleration without a future singularity. By crossing the phantom divide and exhibiting phantom behavior, the Little Rip model explores intriguing aspects of dark energy, such as violation of the null energy condition and potential connections to exotic phenomena. Furthermore, studying the Little Rip model allows for testing and discriminating between different dark energy scenarios, shedding light on the nature of dark energy and its implications for the accelerated expansion. Lastly, the Little Rip model touches upon fundamental aspects of physics, providing insights into the nature of space, time, and energy and offering opportunities to uncover new understanding of the cosmos and fundamental physics. Hence, the Little Rip model represents a crucial subject of investigation in cosmology, addressing outstanding questions, tensions within the cosmological framework, and providing theoretical and observational avenues for further exploration \citep{Frampton_2011}.\\

This paper is organized as follows: Sec. \ref{sec1} gives a brief description of the LR model. In Sec. (\ref{sec2}) we present the method and data used. Then we present the results and discuss them in Sec. (\ref{sec3}). We finish our paper with a conclusion in Sec. (\ref{sec4}).

\section{The Little Rip}\label{sec1}
The Little Rip abrupt event can be envisioned as a scenario in which a Big Rip singularity has been indefinitely delayed. For more information about the LR, we refer the reader to \citep{Stefancic_2005, Nojiri_d71,Nojiri_d72,Frampton_2011}.\\

The equation of state (EoS) characterizing this model is as follows \citep{Bouali_2019}:
\begin{equation}
p_{\text{de}}=-(\rho_{\text{de}} + \beta\sqrt{\rho_{\text{de}}}),
	\label{Eq1}	
\end{equation}
where $\beta$ is a constant characterizing the LR model, with the unit of the square root of energy density, $p_{\text{de}}$ is the DE pressure and $\rho_{\text{de}}$ is the energy density. As a result of the energy momentum tensor's conservation, the energy density $\rho_{\text{de}}$ evolves with the scale factor as 

\begin{equation}
\rho_{\text{de}}=\rho_{\text{de,0}}\left[ \frac{3\beta}{2\sqrt{\rho_{\text{de0}}}}\ln(a)+1\right] ^{2},
	\label{Eq2}	
\end{equation}
where $\rho_{\text{de,0}}$ is the current value of the DE density.\\

The EoS parameter $w_{\text{de}}$ is then
\begin{equation}
w_{\text{de}} = -1-\left( \frac{\beta}{\sqrt{\rho_{\text{de,0}}}+\frac{3}{2}\beta\ln(a)}\right) .
	\label{eos}	
\end{equation}
The value of the EoS parameter, $w_{\text{de}}$, is clearly dependent on the sign of $\beta$. For positive values of $\beta$, we get $w_{\text{de}}<-1$ and the model describes a phantom DE. For negative values of $\beta$,   $w_{\text{de}}>-1$, the model describes a quintessence DE, and deviates from $\Lambda$CDM at $a=1$, by the equation of state $w_{\text{de,0}} = -1-\left( \frac{\beta}{\sqrt{\rho_{\text{de,0}}}}\right)$.\\

On the other hand, the DE equation of state exhibits a singularity associated with the vanishing denominator. More precisely, the function  $X(a) = \sqrt{\rho_{\text{de,0}}} + \tfrac{3}{2}\,\beta \ln (a)$ vanishes at  $a_\star = \exp\!\left(-\frac{2 \sqrt{\rho_{\text{de,0}}}}{3\beta}\right)$. The corresponding pole arises in the past (\(a_\star < 1\)) for \(\beta > 0\), and in the future (\(a_\star > 1\)) for \(\beta < 0\). For either sign of \(\beta\), one finds a phantom regime (\(w_{\text{de}}  < -1\)) for \(a > a_\star\), while a quintessence-like regime (\(w_{\text{de}}  > -1\)) holds for \(a < a_\star\).  This singularity reflects the non-monotonic behavior of the underlying dark-energy density, Eq. (\ref{Eq2}), which forms a parabola in \(\ln(a)\). The minimum of this parabola coincides with the location of the pole in \(w_{\text{de}} (a)\), emphasizing that the dynamics of the EoS are tightly linked to the evolution of the energy density.

By solving the Friedmann equation, the Hubble parameter evolves as follows
\begin{equation}
E^{2}(a) = \Omega_{\text{r,0}} a^{-4} + \Omega_{\text{b,0}} a^{-3} +\Omega_{\text{cdm,0}} a^{-3}+
\Omega_{\text{de,0}} \left[1+\frac{3}{2}\frac{\beta}{\sqrt{\rho_{\text{de,0}}}}\ln(a)\right] ^{2},
	\label{Eq4}
	\end{equation}
where $E(a)=\frac{H(a)}{H_0}$,  $H_0$ is the present value of the Hubble parameter, $\Omega_{\text{r,0}}$, $\Omega_{\text{b,0}}$ and $\Omega_{\text{cdm,0}}$ are the actual fractional energy densities of the radiation, baryon and cold dark matter,  respectively, $\Omega_{\text{de,0}}$ is the present  fractional energy density of dark energy,  with $\Omega_{\text{i,0}}=\rho_{\text{i,0}}/\rho_{\text{cr,0}}$, where $\rho_{\text{cr,0}}=3H_0^2$ is the critical energy density using the units: $8\pi G=c=1$.

\section{Methods and data}\label{sec2}
To obtain the optimal constraints on cosmological parameters, we employ the Markov Chain Monte Carlo (MCMC) method~\citep{Padilla_2021} using the MontePython code~\citep{montepython}, which interfaces with the Boltzmann code CLASS~\citep{class}, where we have constructed our DE fluid. In our analysis, we take into account a 7-dimensional space, which includes the six standard parameters with flat priors, the Hubble constant $H_0$, the density parameters of baryons $\Omega_\text{b,0}h^2$, cold dark matter $\Omega_{\text{cdm,0}}h^2$, the optical depth $\tau$, the scalar spectral index $n_\text{s}$, and the power spectrum amplitude $A_\text{s}$, as well as the additional parameter $\beta$ that defines our DE model. We assume that the  $\beta$-parameter is in the range  [$-0.05$, $0.05$]. To evaluate the convergence quality of our chains, we evaluate the Gelman-Rubin criterion~\citep{R}. After eliminating the burn-in phase points, all cosmological parameters have $R-1<0.01$. On the other hand, to prevent non-adiabatic instabilities in the evolution of perturbations, we employ the Parameterized Post-Friedmann approach~\citep{Fang_2008}.\\

Our statistical analysis is based on various observational data, namely:\\

\textbf{Planck18}: We use the latest CMB measurements from Planck 2018~\citep{aghanim2020planck}  including temperature (TT) and polarization (EE) power spectra at low multipoles $\ell\thicksim$ 2 - 29, as well as the coupled TT-TE-EE at high multipoles $\ell\thicksim$ 30 - 2500 (denoted as CMB in the following).\\

\textbf{BAO}: Baryon Acoustic Oscillation measurements have been used at various redshifts $z$, the Six-degree Field Galaxy Survey (6dFGS) at $z=0.106$~\citep{Beutler_2011}, the Sloan Digital Sky Survey Data Release 7 (SDSS DR7) at $z=0.15$~\citep{Ross_2015}, the latest measurements of the extended Baryon Oscillation Spectroscopic Survey Data Release 16 (eBOSS DR16) at $z=0.38$, $0.51$, $0.7$ and $1.48$~\citep{eBOSS},  as well as,  eBOSS DR16 Ly-$\alpha$-Ly-$\alpha$ and Ly-$\alpha$-quasar sample, where both are at $z=2.33$ \citep{BAO_LY}  (denoted as BAO in the following).\\

\textbf{PantheonPlus}: We also use the recent Supernova data from PantheonPlus with 1701 light curves of 1550 SNIa for different redshifts ranging from z = 0.001 to z = 2.26~\citep{PP} (denoted as PP in the following).\\

In addition, we use the following recent datasets:\\

\textbf{DESI-25}: We also use the latest BAO measurement from DESI Data Release-2 \citep{D1}. The DESI-DR2 experiment relies on two observing programs. The “bright program'', which comprises various galaxy tracers:  Bright Galaxy Sample (BGS) at $z_{\text{eff}}= 0.295$,  Luminous Red Galaxies (LRGs) at $z_{\text{eff}}= 0.510$, $0.706$ and $0.922$, Emission Line Galaxies (ELGs) at  $z_{\text{eff}}= 0.955$ and $1.321$, LRG+ELG at $z_{\text{eff}}= 1.321$. And the “dark program'', which comprises quasars (QSOs) at  $z_{\text{eff}}=  1.484$. We also include Lyman-$\alpha$ forest \citep{Lyman}, which offers a data-point at high redshift,  $z_{\text{eff}}= 2.330$.  All of this data, detailed in Table III and Table IV of Ref. \citep{D1}.\\

\textbf{Union3} : We  include the compilation of 2087  SNIa from 24 datasets, samples of Union3 distributed in the redshift range $0.001<z<2.26$ \citep{Union}.\\

 \textbf{DES-Y5}:  We also  include the distance modulus measurements of 1635 SNIa distributed in the redshift range $0.1<z<1.13$ obtained by DES-Y5 (Dark Energy Survey Supernova-five years) \citep{DES-Y5}.\\

 \textbf{KV-450+DES-Y1}: Finally, in order to compare the $S_8$ estimate by our model, we use the full likelihood of the weak gravitational lensing measurements from (KiDS+VIKING-450) \citep{KiDS1,KiDS11} in addition to    a Gaussian prior from Dark Energy Survey Year 1 measurements   i.e. $S_8=0.776\pm 0.017$ \citep{DES-Y51,Joudaki_2020}.\\

The best-fit of cosmological parameters is estimated by minimizing the chi-square total, $\chi^2_{\text{tot}}$, 
\begin{equation}
\chi^2_{\text{tot}}=\chi^2_{\text{CMB}}+\chi^2_{\text{BAO}}+\chi^2_{\text{SNIa}},
\end{equation}
or by maximizing the Likelihood, $\mathcal{L}_{\text{tot}}$ 
\begin{equation}
\mathcal{L}_{\text{tot}}=\exp{(-\chi^2_{\text{tot}}/2)}.
\end{equation}

To compare the LR model with the base model, $\Lambda$CDM, we calculate the value of $\Delta \chi^2=\chi^2_{\text{LR}}-\chi^2_{\Lambda\text{CDM}}$. In fact, a smaller $\chi^2$ means that the model is better adapted to the observation data. Therefore, a positive (negative) value of $\Delta \chi^2$ indicates that $\Lambda$CDM (LR) has a high quality of fit. However, a higher number of parameters tends to lead to a smaller $\chi^2$. In these case, the $\Delta\chi^2$ is not an appropriate way of comparing models.\\
To confidently predict the goodness of fit between the $\Lambda$CDM and LR models, we use two other statistical tools, the Akaike Information Criteria (AIC) \citep{AIC} and Bayesian analysis of the evidence \citep{evii}. The AIC penalizes models with  extra free parameters, as it depends on the number of parameters, $N_\text{p}$ and the logarithmic of the likelihood, $\mathcal{L}$, where
\begin{equation}
\text{AIC}=-2\ln{(\mathcal{L}_{\mathrm{max}})}+2N_\text{p}.
\end{equation}
The model with the smallest AIC value gives the best fit to the dataset. To compare the LR model with the reference model, $\Lambda$CDM, we calculate $\Delta \text{AIC}=\text{AIC}_{\mathrm{LR}}-\text{AIC}_{\mathrm{\Lambda CDM}}$ and we apply the same rule of $\Delta\chi^2$. We also perform a Bayesian analysis of the evidence, which takes into account the prior information. To calculate Bayes factors ($\mathcal{B}$), we use the MCEvidence\footnote{https://github.com/yabebalFantaye/MCEvidence.} package \citep{evi}, which calculates this factor for each model directly from the MCMC chains created by MontePython. In this work, we have calculated
\begin{equation}
\ln{(\mathcal{B})}= \ln{(\mathcal{P}_{\text{LR}})}-\ln{(\mathcal{P}_{\Lambda\text{CDM}})}.
\end{equation}
$\mathcal{P}_\text{i}$ is the normalization factor in Bayes' theorem \citep{bay} called Bayesian evidence. A positive (negative) value of $\ln{(\mathcal{B})}$ shows that the LR ($\Lambda$CDM) model is preferred\footnote{According to the revised Jeffreys scale \citep{evi}, the selection rules of $\ln{(\mathcal{B})}$, is as follows: for 0$\leqslant\mid\ln{(\mathcal{B})}\mid<$1  means that the  evidence is insufficient, for 1$\leqslant\mid\ln{(\mathcal{B})}\mid<$3, indicates  that there is a positive evidence, for 3$\leqslant\mid\ln{(\mathcal{B})}\mid<$5  indicates  a solid evidence and for  $\mid\ln{(\mathcal{B})}\mid\geqslant$5  indicates a very solid evidence.}. 
\section{Results and discussions}\label{sec3}
\begin{figure*}
\centering
\includegraphics[scale=0.4]{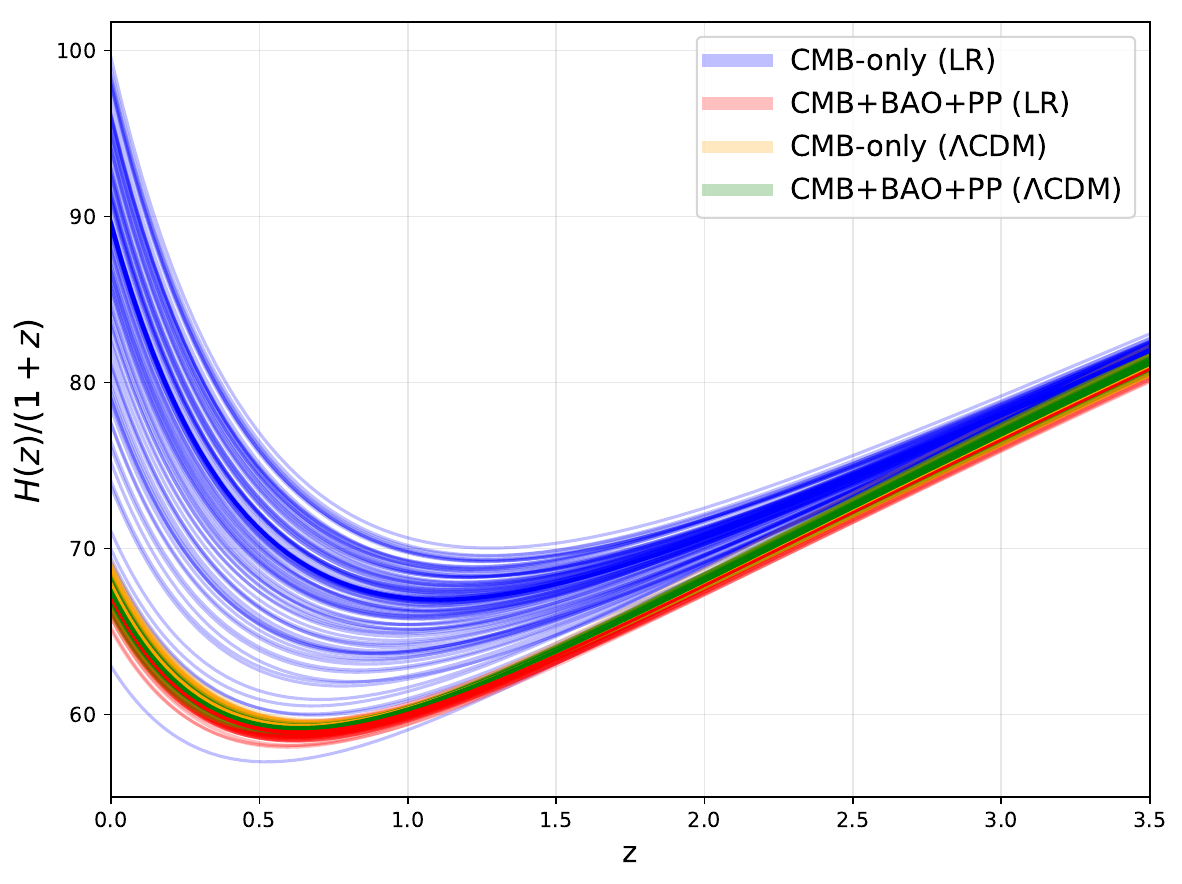}
\includegraphics[scale=0.4]{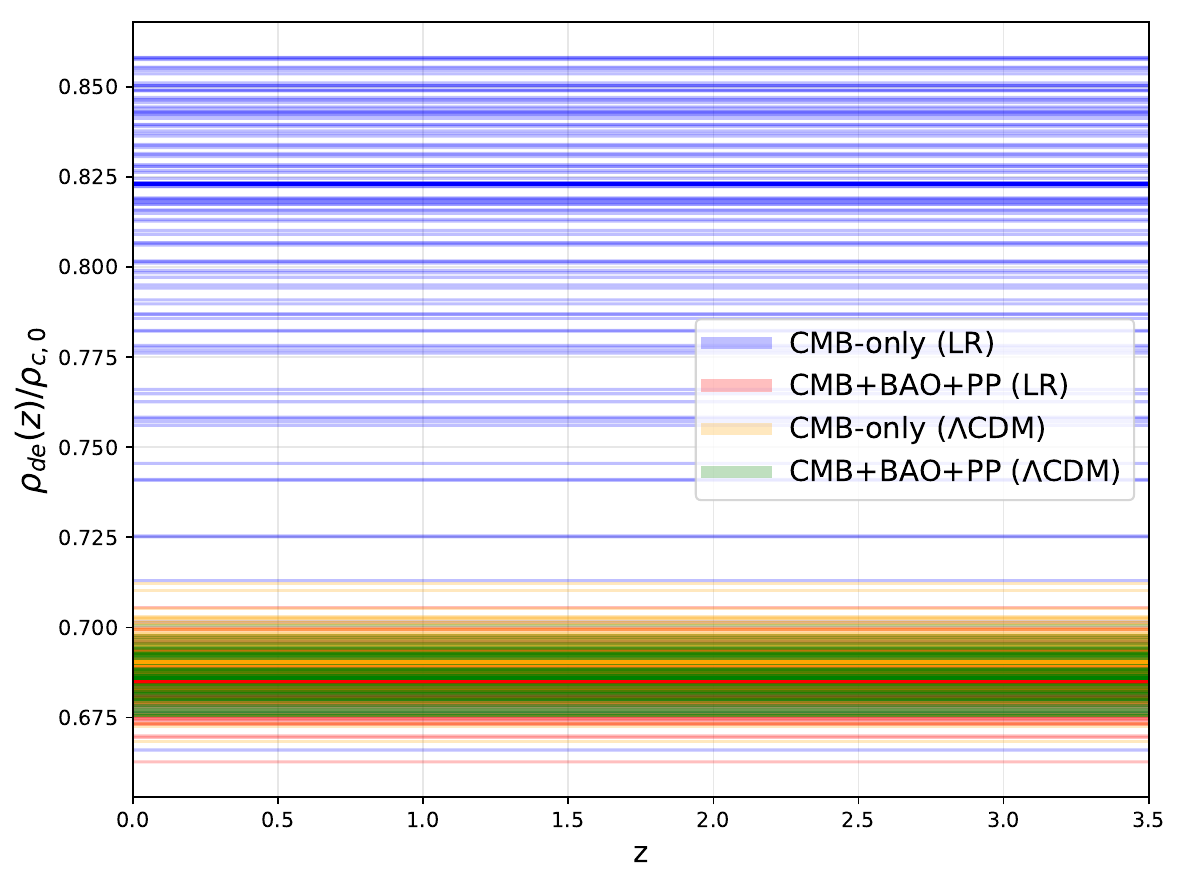}
\includegraphics[scale=0.4]{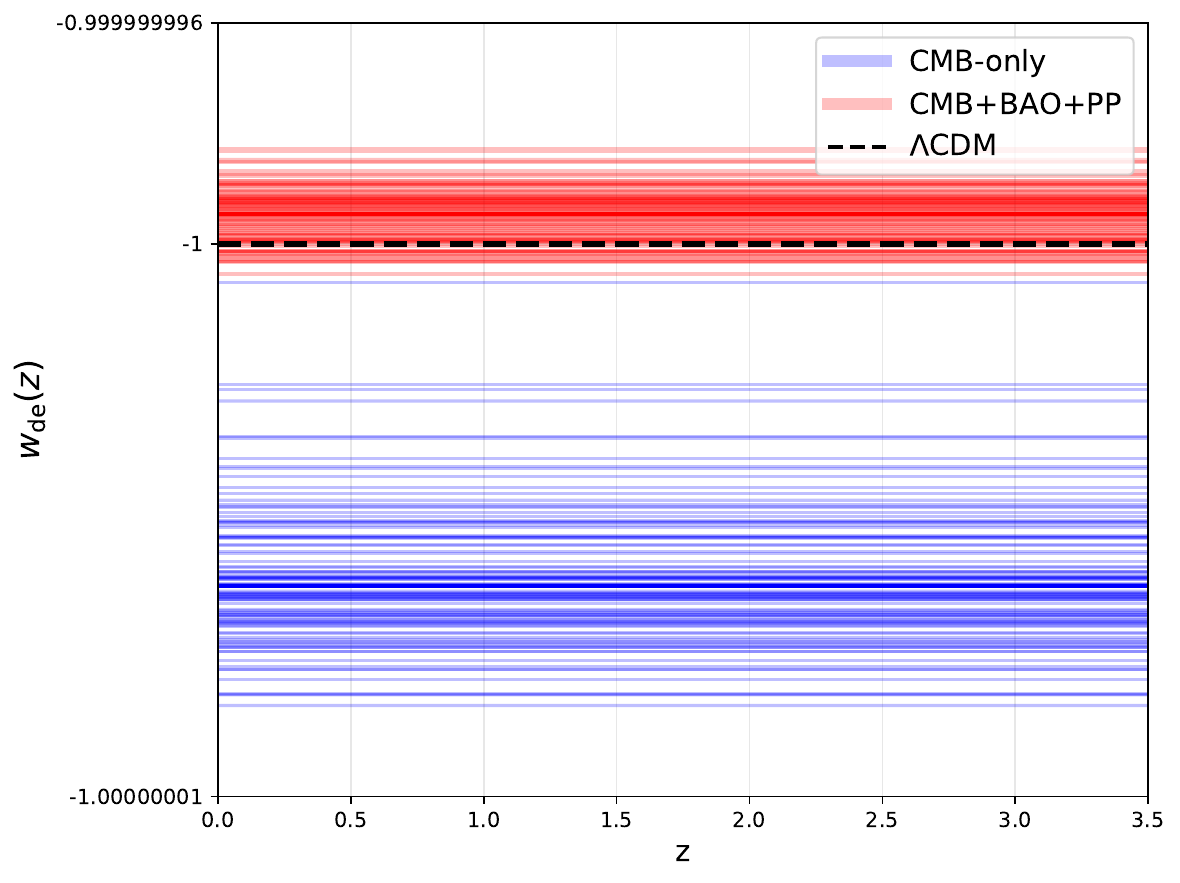}
\caption{Functional posterior of $H(z)/(1+z)$,  $\rho_{\text{de}}(z)/\rho_{\text{c,0}}$ and  $w_{\text{de}}(z)$ plotted for the CMB-only and CMB+BAO+PP datasets for LR and $\Lambda$CDM models.}.
\label{saf}
\end{figure*}

\begin{figure}
\centering
\includegraphics[width=8cm,height=6cm]{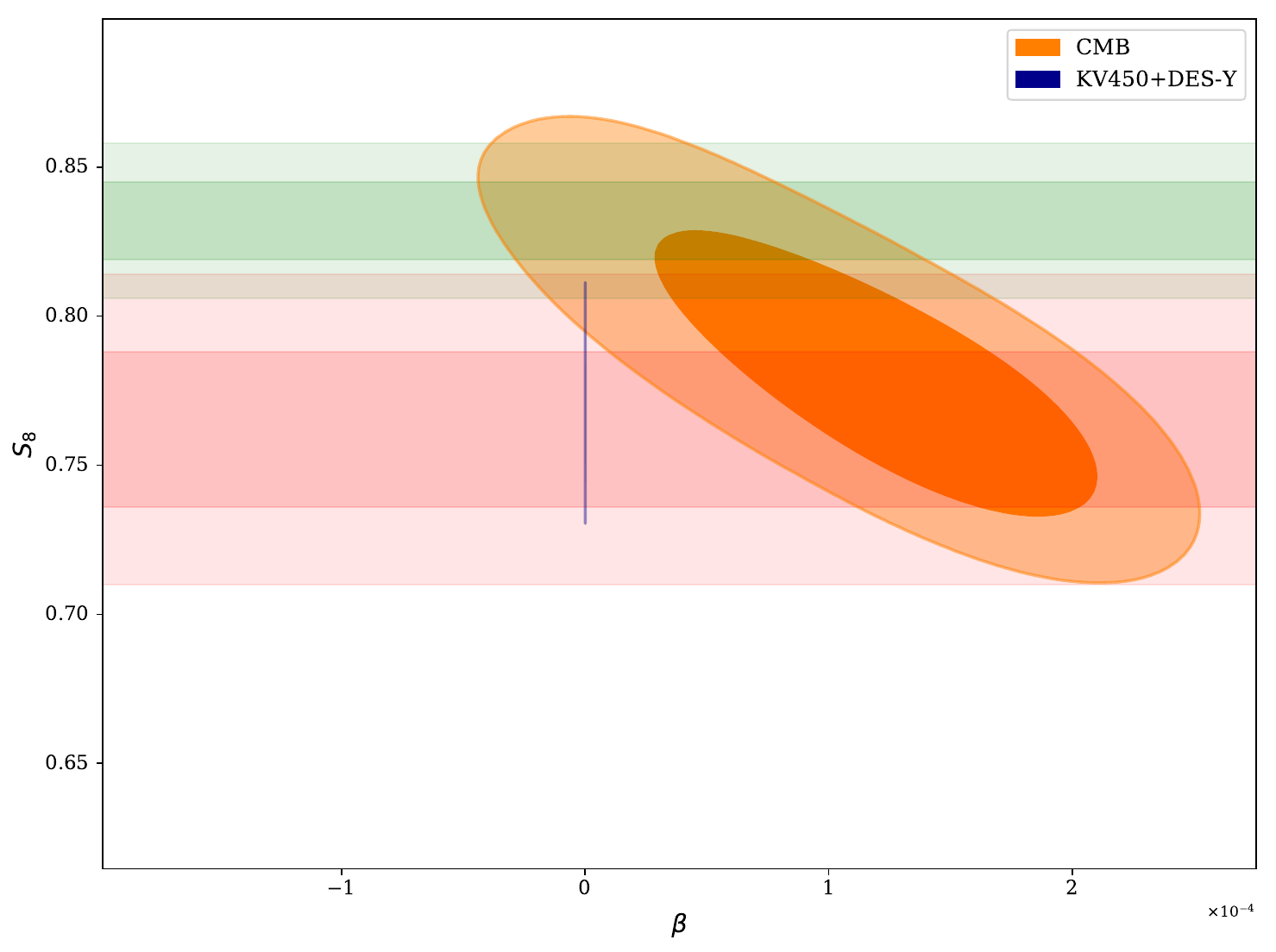} 
\caption{ 2D marginalized contours at 68\% and 95\% confidence levels on $S_8$ and $\beta$ ( The parameter $\beta$ is expressed in units of [$\sqrt{\rho}$], where $\rho$ is the energy density) for the LR model, plotted for CMB (in orange) and KV450+DES-Y1 (in blue). In addition, we present the value obtained by the combination of KV450  and DES-Y1 \citep{Joudaki_2020} (in red) and the Planck18 measurement \citep{aghanim2020planck} (in green).}
\label{s8}
\end{figure}

\begin{figure*}
\centering
\includegraphics[scale=0.4]{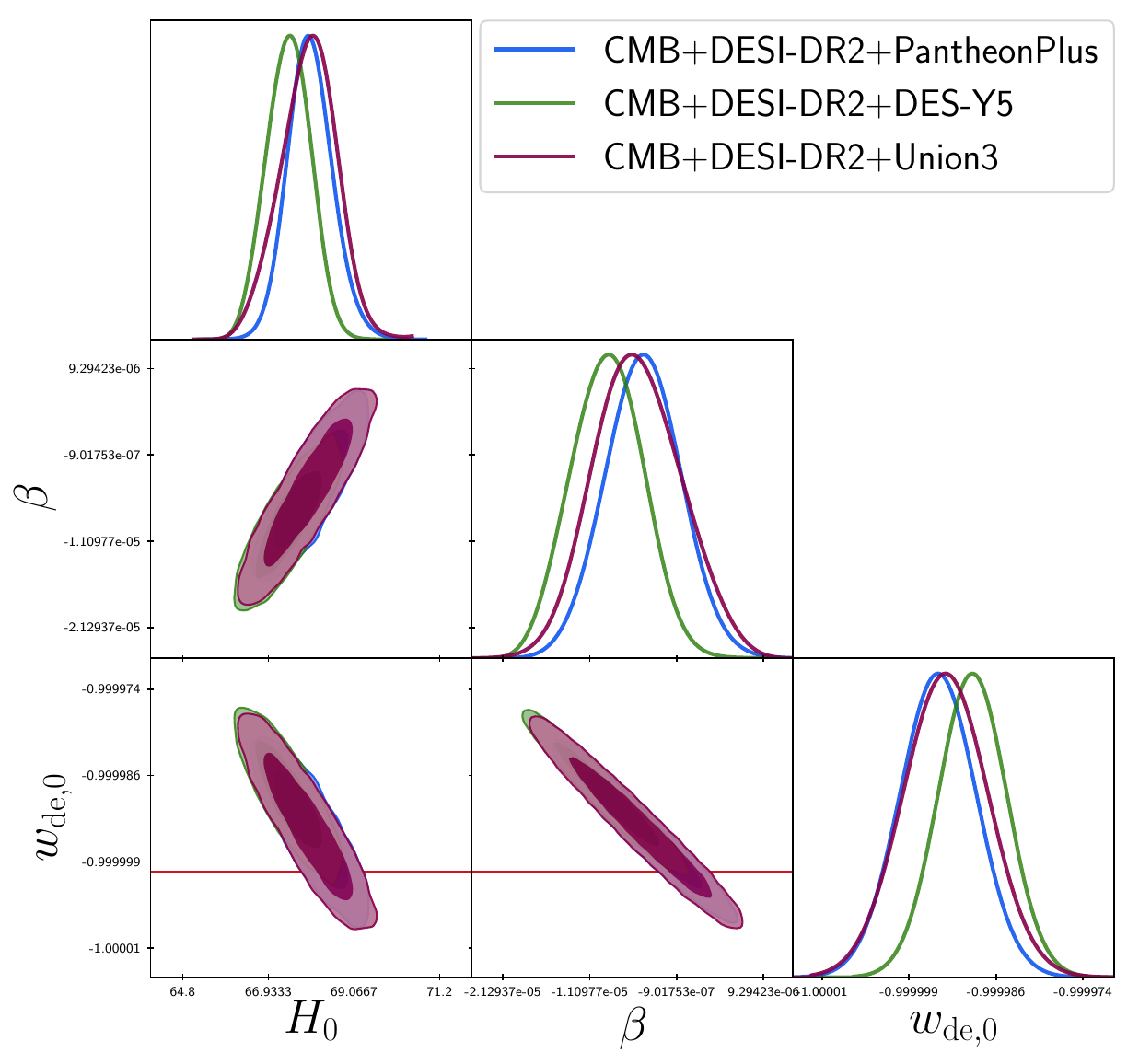}
\includegraphics[scale=0.4]{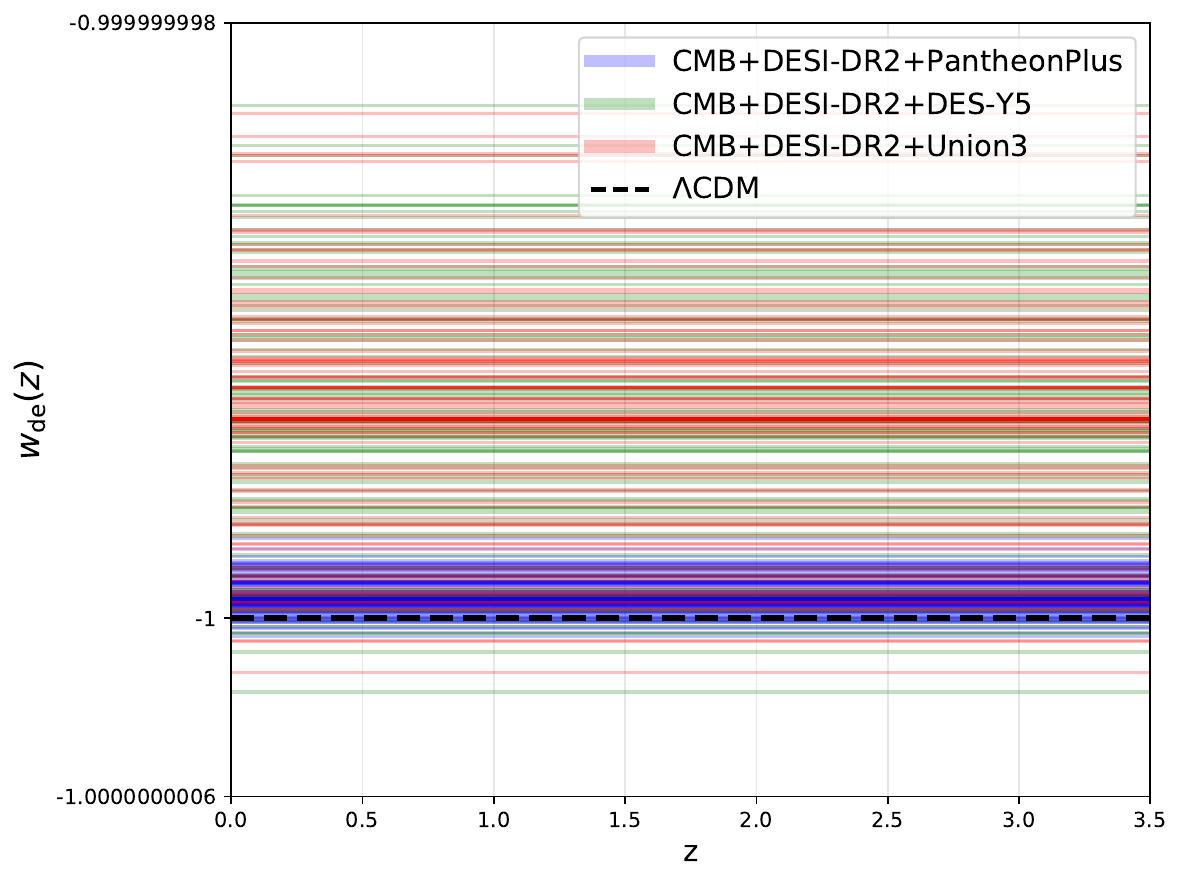}
\caption{The 1D posteriori distributions and 2D marginalized contours at $1\sigma$ and $2\sigma$ for the LR model, using CMB+DESI-DR2+Union3,CMB+DESI-DR2+DES-Y5 and CMB+DESI-DR2+PantheonPlus (left panel), where the parameter $\beta$ is expressed in units of [$\sqrt{\rho}$] and $H_0$ in [km s$^{-1}$ Mpc$^{-1}$]. The right panel shows the functional posterior of $w_{\text{de}}(z)$.}
\label{sa}
\end{figure*}

\begin{figure*}
\centering
\includegraphics[width=17cm,height=6cm]{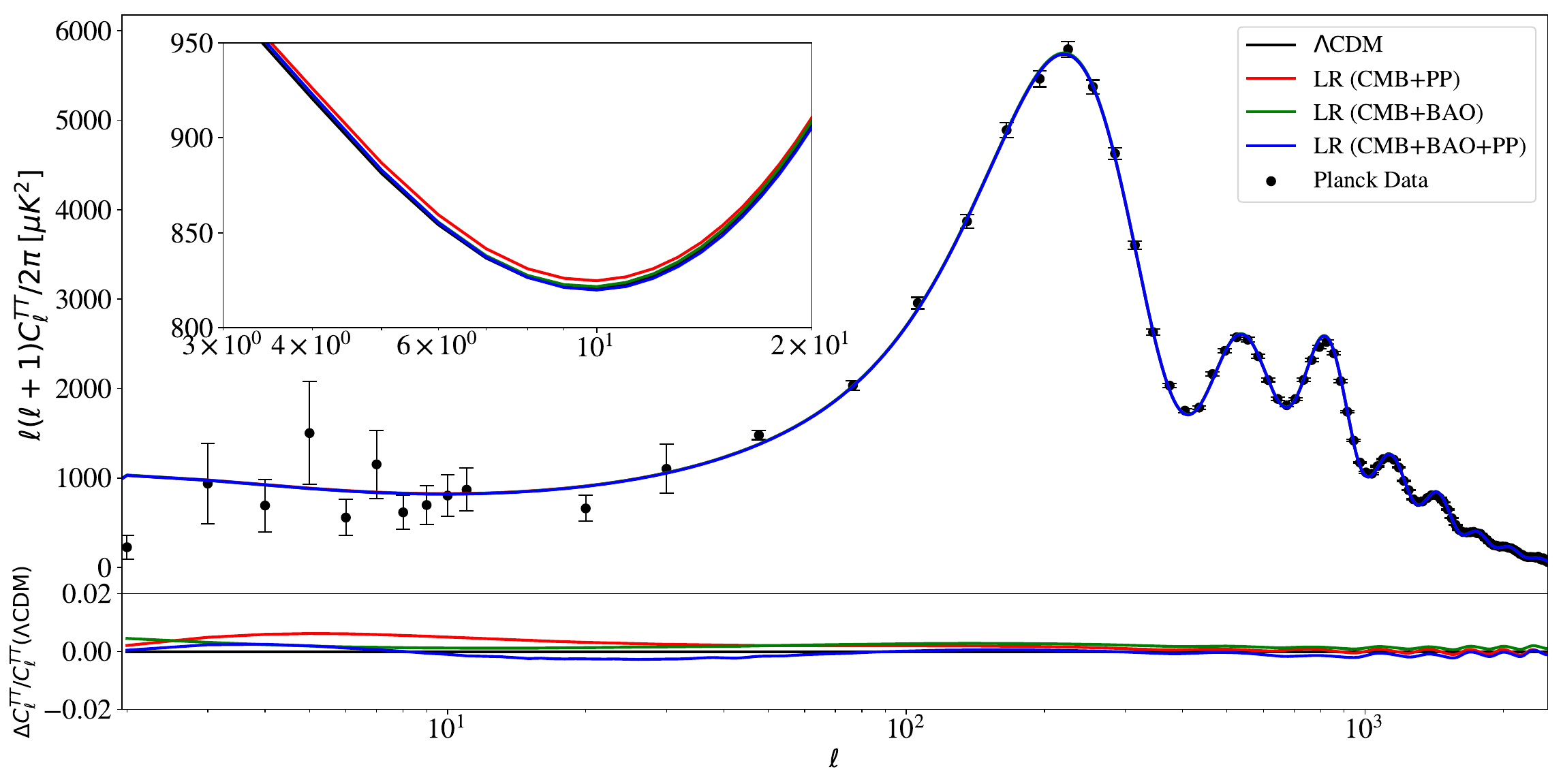} \hfill
\includegraphics[width=11cm,height=6cm]{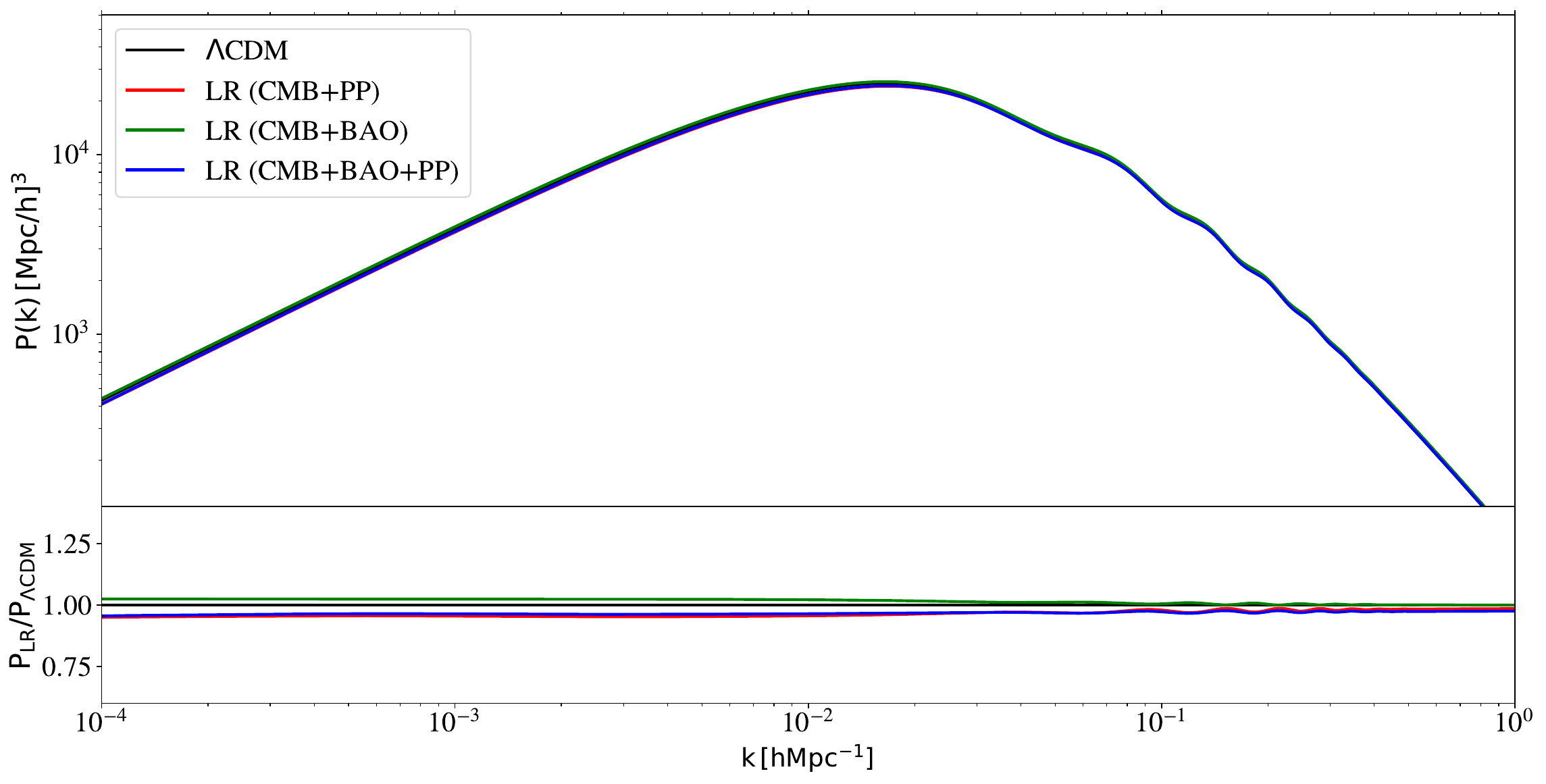}
\caption{The upper panel shows the CMB temperature power spectrum $C_{\ell}^{TT}$ and $\Delta C_{\ell}^{TT}/C_{\ell}^{TT}(\Lambda \text{CDM})$ . The lower panel represents the matter power spectrum $P(k)$ and $P(k)/P_{\Lambda \text{CDM}}(k)$, using the results obtained in Table (\ref{T2}).}
\label{CM}
\end{figure*}

In this section, we discuss the results obtained using the statistical method and datasets described in Sec. (\ref{sec2}). To constrain the $\Lambda$CDM and LR models in the early Universe, we first consider the latest measurements of CMB, then we combine them with BAO and PantheonPlus datasets from the late Universe. In Table (\ref{T2}), we present the 68\% confidence level (C.L.) constraints on all cosmological parameters of the $\Lambda$CDM and LR models obtained using different combinations of datasets as well as the tension level between the value of $H_0$ estimated by both models and the value measured by SH0ES.\\
We constrain the LR model using only CMB data, we obtain a positive value of $\beta=(0.1328_{-0.044}^{+0.072})\times10{^{-3}}$. According to Eq. (\ref{eos}), we obtain, $w_{\text{de}}\sim-1$.  We can deduce that the LR model slightly directed towards a phantom region, when constrained by high redshift data. We also obtain a 95\% C.L. lower limit of $H_0$, i.e. $H_0>72.86$ km s$^{-1}$ Mpc$^{-1}$. The tension with $H_0^{\text{SH0ES}}$ is less than $1\sigma$. For $\Lambda$CDM, we obtain a $H_0$ value around $67$ km s$^{-1}$ Mpc$^{-1}$ and a tension higher than $4\sigma$.  By combining early (CMB) and late (PantheonPlus) data, we observe that the value of $H_0$ decreases significantly to  $66.87_{-0.91}^{+0.87}$  km s$^{-1}$ Mpc$^{-1}$ and the value of $\beta$ becomes negative. The Hubble  tension increases to $4.46\sigma$ for the LR model, and to $5.02\sigma$ for $\Lambda$CDM.  When we combined the CMB data with the BAO dataset, we find  $\beta=(1.49_{-0.11}^{+9.9})\times10{^{-6}}$, we also see a sharp decrease in the mean value of $H_0$ ($H_0= 68.11_{-1.4}^{+1.3}$ km s$^{-1}$ Mpc$^{-1}$) compared to the value obtained by CMB alone and the $H_0$ tension decrease to $\sim 2.8\sigma$, compared to the tension obtained by CMB+PP, while for $\Lambda$CDM the tension remains above $4\sigma$. Adding the supernova measurements from PantheonPlus to CMB+BAO, we get a negative value for $\beta$, $\beta=(-6.06_{-5.3}^{+5.1})\times10{^{-6}}$.  This means that the LR model is slightly directed towards a quintessence region and the value of $H_0$ to a low value ($H_0=67.12_{-0.76}^{+0.72}$ km s$^{-1}$ Mpc$^{-1}$), which is in tension with $H_0^{\text{SH0ES}}$ at $4.5\sigma$. This tension is slightly less than the tension obtained for the $\Lambda$CDM model, which reached $5\sigma$.  At the end of this part, the LR model succeeded in reducing the Hubble tension to less than $3\sigma$ only when we use CMB alone and CMB+BAO, compared to $\Lambda$CDM where Hubble tension remains higher than $4\sigma$, while adding PantheonPlus data, led to a decrease in the value of $H_0$, and, as a result, an increase in Hubble tension.  Furthermore, as shown in Fig. (\ref{T2}), the positive correlation between $\beta$ and $H_0$ may be interpreted as a sign to go beyond the $\Lambda$CDM model in order to overcome or at least to smooth the Hubble tension, i.e., it reflects the consistency to review the $\Lambda$CDM model.\\
In Fig. (\ref{saf}), we illustrate the functional posterior of $\dot a=H(z)/(1+z)$, $\rho_{\text{de}}(z)/\rho_{\text{c,0}}$ ($\rho_{\text{c,0}}=3H^2_0$ is the critical energy density) and $w_{\text{de}}(z)$, for the CMB only and CMB+BAO+PP datasets using the fgivenx package \citep{fgivenx}. The more the lines frequency increases, the more the probability increases. It is important to note the sensitivity of these three quantities to the sign of $\beta$, despite its low value. For positive value of $\beta$, $H(z)$ increases,  while, for its negative value, $H(z)$  approaches that of $\Lambda$CDM. We also note that the extremely low $\beta$ value led to a less dynamic evolution of the equation of state parameter, $w_{\text{de}}(z)$, and the dark energy density, $\rho_{\text{de}}(z)/\rho_{\text{c,0}}$. \\
On the other hand, fitting the LR model by KV450 and DES-Y1  combined, we get  $\beta=(3.116_{-1.1}^{+1.3})\times10{^{-8}}$ and $S_8=0.7702_{-0.015}^{+0.017}$. This value is in agreement with that obtained by KV450 and DES-Y1  combined (i.e.  $S_8 = 0.762^{+0.025}_{-0.024}$), assuming the $\Lambda$CDM model, and is in tension at about $2.88\sigma$ compared to Planck18 (i.e.  $S_8 = 0.832 \pm 0.013$). By fitting our model with CMB only, we obtain $S_8 =0.7765_{-0.034}^{+0.021}$, this value is consistent with the value obtained by KV450+DES-Y1, at $0.05\sigma$ level, (see Fig. (\ref{s8})). \\
In Table (\ref{R}), we show the mean$\pm1\sigma$ of the cosmological parameters for the  $\Lambda$CDM and  LR models, using CMB+DESI-DR2+Union3, CMB+DESI-DR2+PantheonPlus and CMB+DESI-DR2+DES-Y5 datasets. We combine the DESI-DR2 data with the CMB and recent SNIa data, we obtain a negative value for $\beta$, which means that these combinations prefer $w_{\text{de,0}}>-1$, and the model describes a quintessence dark energy. In Fig. (\ref{sa}), we show the 1D posterior distributions and the 2D marginalized contours at $1\sigma$ and $2\sigma$ for the LR model. and the functional posterior of the equation of state parameter, $w_{\text{de}}$. We observe in both figures that the negative value of $\beta$ and quintessence-like regime of dark energy are the most probable results when combining DESI-DR2 data with recent SNIa measurements and CMB data.\\
To compare between the $\Lambda$CDM and LR models, we use the $\Delta \chi^2_{\text{tot}}$, $\Delta \text{AIC}$  and $\ln{(\mathcal{B})}$ criteria, shown in tables (\ref{aic}) and (\ref{R}). We note that for the majority of datasets used, the values of $\Delta \chi^2_{\text{tot}}$ and  $\Delta\text{AIC}$  are very low, so no statistical conclusions can be deduced.  However, for the CMB and CMB+DESI-DR2+DES-Y5 data, the difference between  $\Lambda$CDM and LR slightly increases, we obtain \{$\Delta \chi^2_{\text{tot}}=-4$ , $\Delta\text{AIC}=-2$\} and \{$\Delta \chi^2_{\text{tot}}=-4.898$ , $\Delta\text{AIC}=-2.898$\}, respectively for CMB and CMB+DESI-DR2+DES-Y5, which indicates a positive preference for the LR model.
On the other hand, the sign of $\ln{(\mathcal{B})}$ shows a very solid preference for the LR model over $\Lambda$CDM, only for CMB data, $\ln{(\mathcal{B})}=+6.78$ and $\mid\ln{(\mathcal{B})}\mid\geqslant$5. For all other datasets, we obtain a negative value for $\ln{(\mathcal{B})}$ and $\mid\ln{(\mathcal{B})}\mid\geqslant$5, indicating a very solid preference for the  $\Lambda$CDM model over the LR model. This result reflects the different penalty structures of the two approaches: $\Delta$AIC assigns only a small penalty to one extra parameter, whereas Bayesian evidence includes a prior-volume penalty. Consequently, given our choice of a uniform prior for $\beta$, $\beta\in$ [$-0.05$, $0.05$], the additional parameter is only weakly supported. This can also be explained by the fact that the BAO/DESI-DR2 data and supernova measurements are not fully consistent when combined with CMB, resulting in a value close to 0 for $\beta$, i.e. the model reduces to $\Lambda$CDM.\\
After studying the global fits of the $\Lambda$CDM and LR models to the data, we study the effect of the different values of $\beta$ obtained from MCMC analyses on CMB temperature power spectrum, $C_{\ell}^{TT}$ and the matter power spectrum $P(k)$, as shown in Fig. (\ref{CM}). We notice that the effect of the LR model on the CMB temperature power spectrum is negligible and significant only for  $\beta=(-5.82_{-5.9}^{+5.4})\times10{^{-6}}$, obtained by CMB+PP data, particularly
on a large scale $\ell < 50$. In the bottom panel of Fig. (\ref{CM}), we show the current matter power spectrum $P(k)$, for the $\Lambda$CDM and the LR models. We note that our model is indistinguishable from $\Lambda$CDM, for all value of $\beta$. Although there is a slight difference between LR and $\Lambda$CDM, which is observed for the negative value of $\beta$,  obtained by CMB+PP and CMB+BAO+PP.
\begin{table*}
\centering
{\caption{Summary of the mean$\pm1\sigma$ of the cosmological parameters for the  $\Lambda$CDM and  LR models, using CMB, CMB+PP, CMB+BAO and  CMB+BAO+PP datasets as well as the level of tension  between the estimated value of $H_0$ and the SH0ES measurements, for all combinations of datasets studied in this work. To quantify this tension, we use the following expression: $T=\mid H_0-H_0^{\text{R21}} \mid/\sqrt{\sigma^2(H_0)+\sigma^2(H_0^{\text{R21}} )}$ \citep{ten}, The values of $\beta$ are given in units of [$\sqrt{\rho}$], where $\rho$ is the energy density.}\label{T2}}
\scalebox{0.75}{
\begin{tabular}{c|cc|cc|cc|cc}
\hline
\hline
\multicolumn{1}{c|}{Data} & \multicolumn{2}{c|}{CMB} & \multicolumn{2}{c|}{CMB+PP}&\multicolumn{2}{c|}{CMB+BAO}&\multicolumn{2}{c}{CMB+BAO+PP}\\
\hline
\multicolumn{1}{c|}{Model} & \multicolumn{1}{c}{$\Lambda$CDM} & \multicolumn{1}{c|}{LR}& \multicolumn{1}{c}{$\Lambda$CDM} & \multicolumn{1}{c|}{LR} & \multicolumn{1}{c}{$\Lambda$CDM} & \multicolumn{1}{c|}{LR}& \multicolumn{1}{c}{$\Lambda$CDM} & \multicolumn{1}{c}{LR}\\
\hline
$100\Omega_\text{b,0}h^2$   & $2.237_{-0.017}^{+0.016}$  & $2.241\pm{0.016}$ &$2.231\pm{0.015}$& $2.236\pm{0.016}$
 & $2.238_{-0.015}^{+0.014}$ & $2.237\pm{0.015}$ & $2.238_{-0.015}^{+0.014}$ & $2.24_{-0.016}^{+0.015}$  \\[0.1cm]
$\Omega_\text{cdm,0}h^2$    & $0.1201_{-0.0014}^{+0.0015}$  & $0.1197\pm{0.0015}$&$0.1207\pm{0.0013}$&  $0.1202_{-0.0015}^{+0.0014}$ & $0.1199\pm{0.0011}$  & $0.12\pm{0.0013}$ & $0.1199\pm{0.001}$ & $0.1195\pm{0.0012}$  \\[0.1cm]
$\tau_{\mathrm{reio}}$   & $0.05425_{-0.0083}^{+0.0081}$ &  $0.05423_{-0.0084}^{+0.0083}$ & $0.05399_{-0.0085}^{+0.0079}$&  $0.05404_{-0.0094}^{+0.0084}$ & $0.05429_{-0.0091}^{+0.0082}$ & $0.05414_{-0.008}^{+0.0076}$ & $0.0549\pm{0.0088}$ &  $0.05518\pm{0.0085}$\\[0.1cm]
$n_{\mathrm{s} }$    &  $0.9649\pm{0.0046}$&  $0.966_{-0.0049}^{+0.0047}$  & $0.9637_{-0.0046}^{+0.0043}$ &  $0.9649_{-0.0047}^{+0.0046}$  & $0.9656_{-0.0038}^{+0.0041}$& $0.9654\pm{0.0042}$ & $0.965\pm{0.004}$ & $0.9665\pm{0.0043}$ \\[0.1cm]
$\ln{(10^{10}A_{\mathrm{s} })}$   & $3.044\pm{0.017}$  & $3.044\pm{0.017}$ & $3.045_{-0.018}^{+0.016}$&  $3.044_{-0.019}^{+0.017}$ & $3.044_{-0.018}^{+0.017}$ & $3.044\pm{0.016}$ & $3.045_{-0.018}^{+0.017}$ & $3.045\pm{0.017}$  \\[0.1cm]
$\beta$   & -  & $(0.1328_{-0.044}^{+0.072})\times10{^{-3}}$  &-&  $(-5.82_{-5.9}^{+5.4})\times10{^{-6}}$& - & $(1.49_{-0.11}^{+9.9})\times10{^{-6}}$ & - & $(-6.06_{-5.3}^{+5.1})\times10{^{-6}}$ \\[0.1cm]
\hline
$\Omega_{\textrm{m}}$    & $0.3098_{-0.0088}^{+0.0087}$&  $<0.267$ (95\%) & $0.3195_{-0.0083}^{+0.0081}$&  $0.3189_{-0.0099}^{+0.0096}$ & $0.3084_{-0.0063}^{+0.0064}$  & $0.3072\pm{0.012}$& $0.3141_{-0.0063}^{+0.0061}$& $0.3151_{-0.0075}^{+0.0072}$  \\[0.1cm]
$\sigma_8$   & $0.824_{-0.0076}^{+0.008}$ &  $0.9967_{-0.04}^{+0.082}$&$0.8128_{-0.008}^{+0.0077}$& $0.8157_{-0.014}^{+0.013}$ & $0.8233_{-0.0083}^{+0.0077}$  &  $0.826_{-0.02}^{+0.019}$ & $0.8104_{-0.008}^{+0.0076}$  & $0.8139\pm{0.013}$ \\[0.1cm] 
$M_\mathrm{B}$ [mag]  & -  & - &$-19.45_{-0.017}^{+0.016}$ &  $-19.45\pm{0.023}$ & -& - & $-19.44\pm{0.013}$  & $-19.44_{-0.019}^{+0.018}$   \\[0.1cm]
$H_0$ [km s$^{-1}$ Mpc$^{-1}$] & $67.83_{-0.64}^{+0.63}$  & $>72.86$ (95\%) &$67.06_{-0.6}^{+0.58}$ &  $66.87_{-0.91}^{+0.87}$ & $67.93_{-0.48}^{+0.46}$  & $68.11_{-1.4}^{+1.3}$ & $67.45_{-0.45}^{+0.46}$  & $67.12_{-0.76}^{+0.72}$   \\[0.1cm]
$S_8$  &   $0.8373_{-0.016}^{+0.017}$   & $0.7765_{-0.034}^{+0.021}$ & $0.8389_{-0.016}^{+0.015}$  & $0.8408_{-0.016}^{+0.015}$ & $0.8347_{-0.013}^{+0.013}$ & $0.8354_{-0.013}^{+0.013}$ &  $0.8293\pm{0.012}$   & $0.834_{-0.013}^{+0.013}$   \\[0.1cm]

\hline    
H$_0$ tension  &$4.26\sigma$ &$1.66\sigma$ &$5.02\sigma$& $4.46\sigma$ &$4.46\sigma$ &$2.8\sigma$&$5\sigma$ &$4.5\sigma$
 \\[0.1cm]
\hline
\hline

\end{tabular}
}
\end{table*}
\begin{figure*}
\centering
\includegraphics[width=8.2cm,height=7cm]{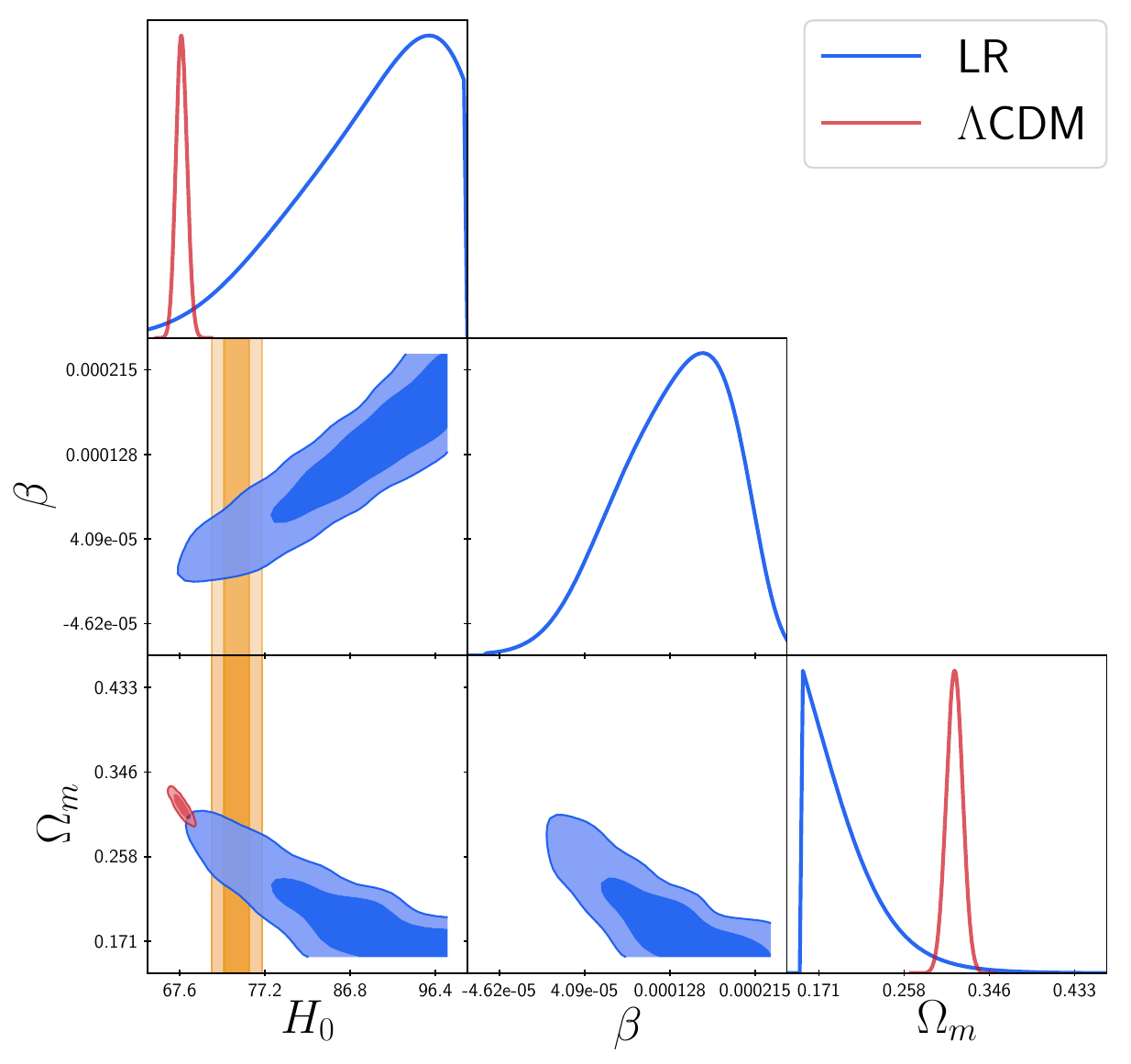} \hfill
\includegraphics[width=8.2cm,height=7cm]{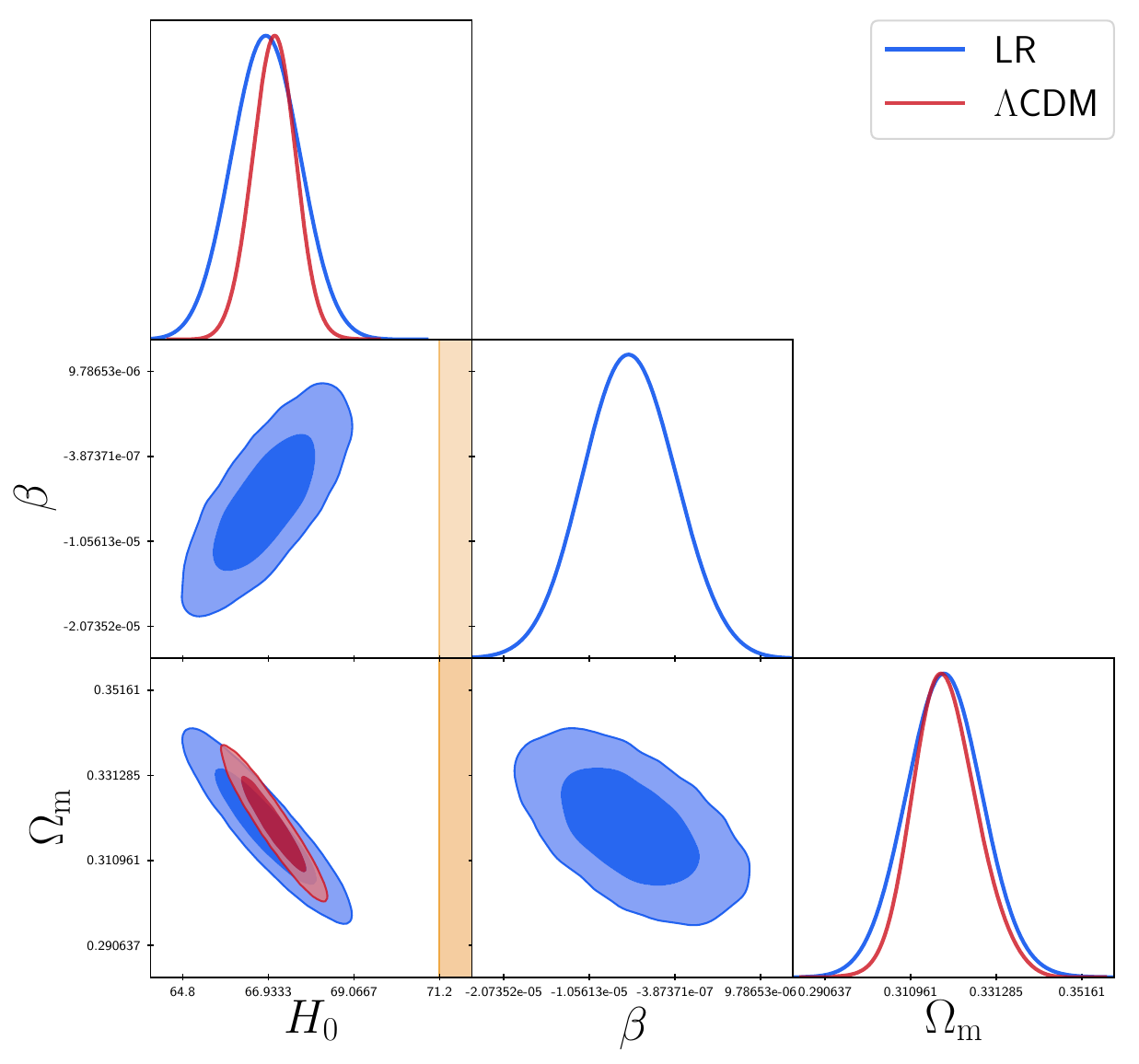}\hfill
\includegraphics[width=8.2cm,height=7cm]{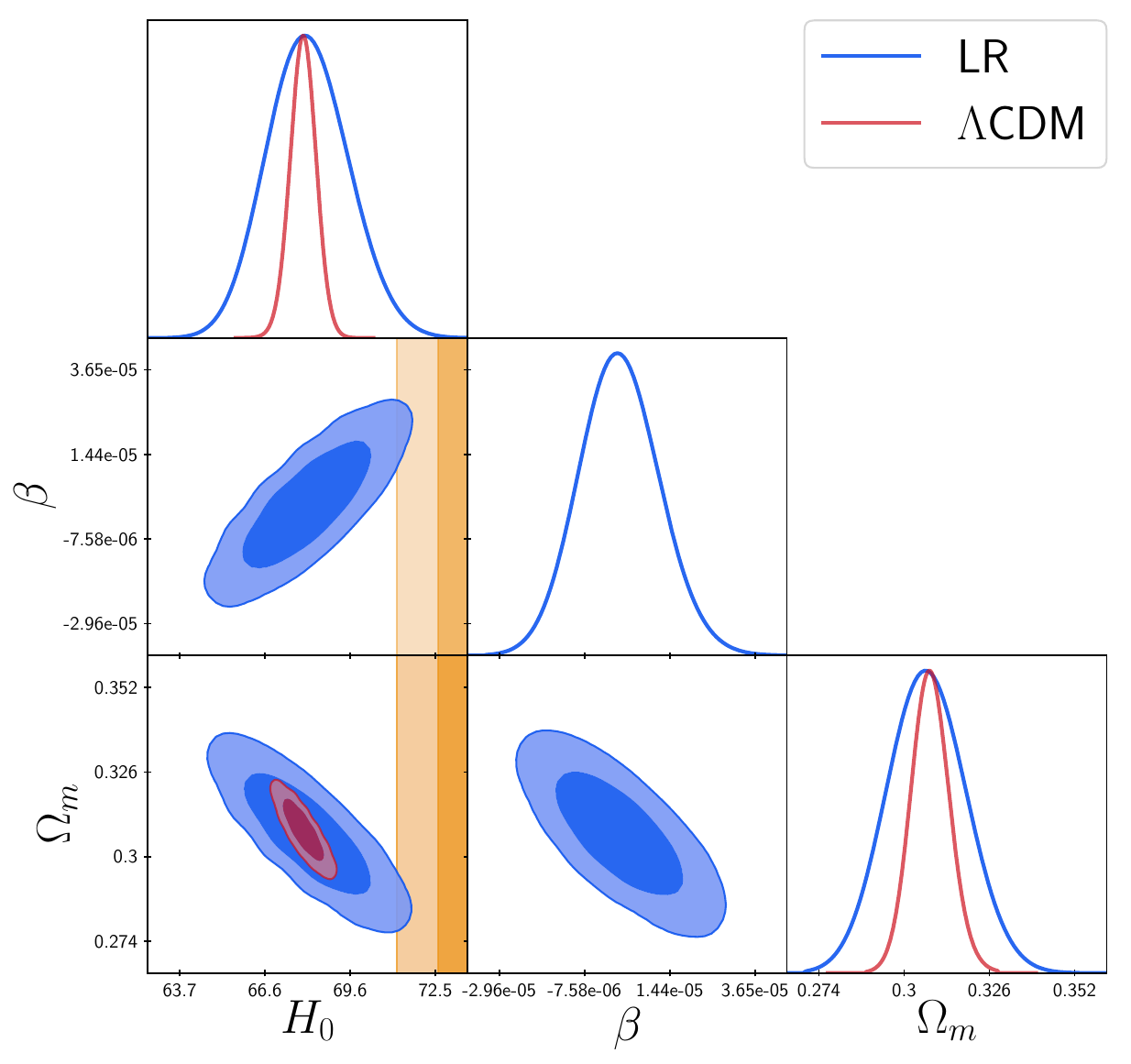}\hfill
\includegraphics[width=8.2cm,height=7cm]{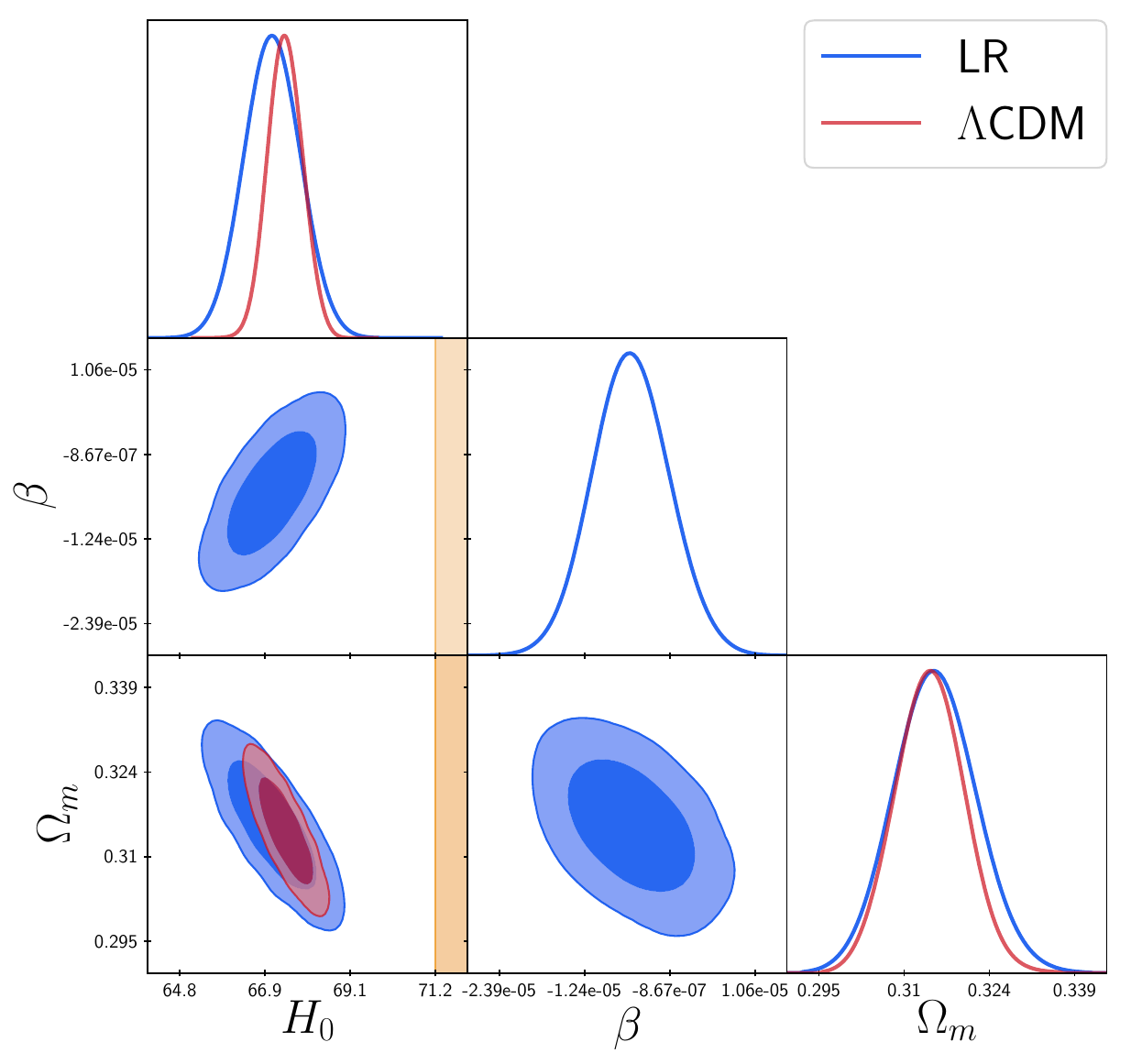} 

\caption{A comparison of 1D posteriori distributions and 2D marginalized contours at $1\sigma$ and $2\sigma$ between $\Lambda$CDM (purple) and LR (blue), for all combinations of datasets: CMB (Top left panel), CMB+PP (Top right panel), CMB+BAO (bottom left panel) and CMB+BAO+PP (bottom right panel). The orange bands represent the local measurements of $H_0$ from SH0ES The parameter $\beta$ is expressed in units of [$\sqrt{\rho}$], where $\rho$ is the energy density and $H_0$ in [km s$^{-1}$ Mpc$^{-1}$]}.
\label{con}
\end{figure*}

\begin{table*}
\centering
{\caption{The  $\chi^2$ per experiment, $\chi^2_{\text{tot}}$, $\Delta \chi^2_{\text{tot}}$, AIC, $\Delta$AIC and  $\ln{(\mathcal{B})}$ for the $\Lambda$CDM and LR models.}\label{aic}}
\scalebox{0.8}{
\begin{tabular}{c|cc|cc|cc|cc}
\hline
\hline
\multicolumn{1}{c|}{Data} & \multicolumn{2}{c|}{CMB} & \multicolumn{2}{c|}{CMB+PP}&\multicolumn{2}{c|}{CMB+BAO}&\multicolumn{2}{c}{CMB+BAO+PP}\\
\hline
\multicolumn{1}{c|}{Model} & \multicolumn{1}{c}{$\Lambda$CDM} & \multicolumn{1}{c|}{LR}& \multicolumn{1}{c}{$\Lambda$CDM} & \multicolumn{1}{c|}{LR} & \multicolumn{1}{c}{$\Lambda$CDM} & \multicolumn{1}{c|}{LR}& \multicolumn{1}{c}{$\Lambda$CDM} & \multicolumn{1}{c}{LR}\\
\hline    

\hline 
Planck\_high-$\ell$\_TTTEEE\_lite      & $ 583.59$ &$580.76$  &$584.102$&$584.48$& $583.65$ & $583.38$ &$585.1$&$ 584.31$
\\[0.1cm]
Planck\_low-$\ell$\_EE    & $  396.04$ &$395.69$   &$396.43$&$395.80$ & $395.93$  & $396.27$ &$395.92$&$ 396.24$
\\[0.1cm]
Planck\_low-$\ell$\_TT    & $23.41$ &$22.59$  &$23.743$&$23.68$ & $23.57$  & $23.53$ &$23.23$&$ 23.23$
\\[0.1cm]
bao\_eboss\_dr16\_gal\_QSO       & - & -   &-&- & $5.82$  & $5.74$ &$ 6.65$&$ 5.84$
\\[0.1cm]
bao\_smallz\_2014      & - &-   &-&- & $1.34$  & $1.22$ &$0.99$&$ 0.99$
\\[0.1cm]
bao\_eBOSS\_DR16\_Lya\_cross\_QSO      & - &-   &-&- & $3.045$  & $3.07$ &$3.22$&$ 3.18$
\\[0.1cm]
bao\_eBOSS\_DR16\_Lya\_auto   & - &-   &-&- & $1.58$  & $1.60$ &$1.71$&$ 1.68$
\\[0.1cm]
PantheonPlus       & - &-  &$1410.59$&$1409.48$ & -  &  - &$1410.79$&$ 1410.04$\\[0.1cm]
\hline
\hline
$\chi^2_{\text{tot}}$    & $1003.06$ &$ 999.06$  &$2414.86$&$2413.45$ & $1014.96$  & $1014.85$ &$2427.65$&$ 2425.53$
\\[0.1cm]
$\Delta  \chi^2_{\text{tot}}$     & $0$ &$-4$  &$0$&$-1.41$ & $0$  & $-0.11$ &$0$&$ -2.12$
\\[0.1cm]
$\text{AIC}$      &$1021.06$ &$1019.06$ &$2434.86$&$2435.45$ & $1032.96$  & $1034.85$ &$2447.65$&$ 2447.53$
\\[0.1cm]
$\Delta  \text{AIC}$       & $0$ &$-2$ &$0$&$+0.59$ & $0$  & $+1.89$ &$0$&$ -0.12$
\\[0.1cm]
\hline
$\ln{(\mathcal{B})}$      & $0$ & $+6.78$ &$0$&$-20.7$ & $0$  & $-6.471$ &$0$&$ -5.663$
\\[0.1cm]
Strength     & $-$ & very solid (LR) &$-$&very solid ($\Lambda$CDM) & $-$  & very solid ($\Lambda$CDM) &$-$& very solid ($\Lambda$CDM)
\\[0.1cm]
\hline 
\hline        
\end{tabular}
}
\end{table*}

\begin{table*}
\centering
{\caption{Summary of the mean$\pm1\sigma$ of the cosmological parameters for the $\Lambda$CDM and LR models, using CMB+DESI-DR2+PantheonPlus, CMB+DESI-DR2+DES-Y5 and  CMB+DESI-DR2+Union3 datasets.  The values of $\beta$ are given in units of [$\sqrt{\rho}$], where $\rho$ is the energy density.\label{R}}}
\scalebox{0.8}{
\begin{tabular}{c|cc|cc|cc}
\hline
\hline
\multicolumn{1}{c|}{Data} &\multicolumn{2}{c|}{CMB+DESI-DR2+PantheonPlus}&\multicolumn{2}{c}{CMB+DESI-DR2+DES-Y5}& \multicolumn{2}{c|}{CMB+DESI-DR2+Union3}\\
\hline
\multicolumn{1}{c|}{Model} & \multicolumn{1}{c}{$\Lambda$CDM} & \multicolumn{1}{c||}{LR}& \multicolumn{1}{c}{$\Lambda$CDM} & \multicolumn{1}{c|}{LR} & \multicolumn{1}{c}{$\Lambda$CDM} & \multicolumn{1}{c|}{LR}\\
\hline
 \hline
$100\Omega_\text{b,0}h^2$      &$2.247_{-0.014}^{+0.013}$ &$2.252^{+0.015}_{-0.014}$ &  $2.245_{-0.014}^{+0.013}$  & $2.253\pm{0.014}$ & $2.248\pm{0.014}$ & $2.251_{-0.015}^{+0.014}$  \\[0.1cm]

$\Omega_\text{cdm,0}h^2$   & $0.1186_{-0.00071}^{+0.00069}$ &$0.118 \pm 0.00088$   &   $0.1188_{-0.00071}^{+0.0007}$  &  $0.1177_{-0.00085}^{+0.0009}$ &  $0.1185\pm{0.00071}$ &  $0.118_{-0.00087}^{+0.00092}$ \\[0.1cm]

$\tau_{\mathrm{reio}}$      &$0.05764_{-0.0079}^{+0.0073}$ &$0.05961^{+0.0077}_{-0.0085}$& $0.05707_{-0.0078}^{+0.0074}$  & $0.06086_{-0.009}^{+0.0077}$ & $0.05784_{-0.0078}^{+0.0074}$ &$0.05991_{-0.0084}^{+0.0077}$ \\[0.1cm]

$n_{\mathrm{s} }$      & $0.9687\pm{0.0036}$& $0.9701^{+0.0037}_{-0.0038}$ & $0.9682_{-0.0036}^{+0.0035}$  &  $0.9708_{-0.0038}^{+0.0037}$   & $0.9688_{-0.0036}^{+0.0037}$ &  $0.9701_{-0.0038}^{+0.004}$ \\[0.1cm]

$\ln{(10^{10}A_{\mathrm{s} })}$   & $3.048_{-0.016}^{+0.015}$ &  $3.051^{+0.015}_{-0.017}$ & $3.047\pm{0.015}$ & $3.053_{-0.018}^{+0.015}$  & $3.049_{-0.016}^{+0.015}$ & $3.052_{-0.017}^{+0.016}$  \\[0.1cm]

$\beta$     &$-$ &$(-4.845_{-5}^{+4.8})\times10{^{-6}}$   & $-$& $(-8.926_{-4.2}^{+4.5})\times10{^{-6}}$  &   $-$ & $(-5.159_{-5.5}^{+6.2})\times10{^{-6}}$ \\[0.1cm]

\hline
$\Omega_{\textrm{m}}$    &  $0.3003_{-0.0041}^{+0.0039}$ & $0.3041^{+0.0056}_{-0.0057}$ &  $0.3016_{-0.0041}^{+0.0039}$  & $0.3085_{-0.0054}^{+0.0051}$ & $0.3001_{-0.0041}^{+0.004}$ &  $0.3046_{-0.0068}^{+0.0061}$ \\[0.1cm]

$\sigma_8$      & $0.8221_{-0.0067}^{+0.0064}$  &$0.8141^{+0.011}_{-0.01}$ &  $0.8244_{-0.0091}^{+0.0092}$ & $0.8077_{-0.0098}^{+0.0097}$&  $0.822_{-0.0066}^{+0.0064}$  & $0.8139_{-0.011}^{+0.012}$ \\[0.1cm]

$S_8$      & $0.8226_{-0.0087}^{+0.008}$ & $0.8195_{-0.0088}^{+0.0093}$ &  $0.8244_{-0.0085}^{+0.0087}$& $0.819_{-0.009}^{+0.0092}$ &  $0.8221_{-0.0087}^{+0.0084}$   & $0.8201_{-0.0092}^{+0.0094}$ \\[0.1cm]
 
$r_{\textrm{s}}$    & $147.4_{-0.2}^{+0.19}$ &$147.5^{+0.21}_{-0.23}$  & $147.6\pm{0.21}$ &$147.6_{-0.22}^{+0.2}$ &  $147.4\pm{0.2}$ &  $147.5\pm{0.21}$\\[0.1cm]
$H_0$ [km s$^{-1}$ Mpc$^{-1}$]   & $68.53_{-0.31}^{+0.32}$ &$67.99^{+0.61}_{-0.66}$ &  $68.44_{-0.32}^{+0.31}$ & $67.44\pm{0.57}$  & $68.55\pm{0.32}$ & $67.93_{-0.71}^{+0.79}$ \\[0.1cm]

\hline
\hline
$w_{\textrm{de,0}}$  & $-1$  &  $-0.999994_{-5.75596\times10{^{-6}}}^{+5.53354\times10{^{-6}}}$&  $-1$  & $-0.999989_{-5.44855\times10{^{-6}}}^{+5.01532\times10{^{-6}}}$  & $-1$  & $0.999994_{-7.54339\times10{^{-6}}}^{+6.62017\times10{^{-6}}}$ \\[0.1cm]

\hline
\hline
$\chi^2_{\text{tot}}$    &$2438.196$ &$2437.424$ &$2674.86$&$2669.962$ & $1053.534$  & $1052.709$ 
\\[0.1cm]
$\Delta \chi^2_{\text{tot}}$     &$0$ &$-0.772$ &$0$&$-4.898$ & $0$  & $-0.825$ 
\\[0.1cm]
$\text{AIC}$      &$2458.196$ &$2459.424$ &$2694.86$&$2691.962$ & $1073.534$  & $1074.709$ 
\\[0.1cm]
$\Delta \text{AIC}$        &$0$ &$+1.228$ &$0$&$-2.898$ & $0$  & $+1.175$ 
\\[0.1cm]
\hline
$\ln{(\mathcal{B})}$      &$0$ &$-9.256$ &$0$&$-6.5705$ & $0$  & $-6.570$ 
\\[0.1cm]
Strength     & $-$&very solid ($\Lambda$CDM) & $-$  & very solid ($\Lambda$CDM) &$-$& very solid ($\Lambda$CDM)
\\[0.1cm]
\hline 
\hline        
\end{tabular}
}
\end{table*}


\section{CONCLUSION}\label{sec4}

In this paper, we presented a study of the $H_0$ and $S_8$ tensions by considering the LR model. This study is based on the fitting of the most recent observation data from CMB, BAO, SNIa and DESI-25 datasets. We also used the most reliable criteria to compare the LR model with the standard model and select the best model, namely the Akaike Information Criteria and the Bayesian analysis of the evidence. Finally, we studied the effect of our model on CMB spectrum. \\

In our initial analysis, we used the CMB and CMB+BAO dataset and showed that for the LR model, the $H_0$ tension is $1.66\sigma$ and $2.8\sigma$, respectively. While for $\Lambda$CDM it remains higher than the $4\sigma$ using the same dataset. Adding Supernova data shows the inability of the LR  model to solve the Hubble tension.  In conclusion to the first part, we have found that, although the LR model reduces the Hubble tension compared to the $\Lambda$CDM model, it does not achieve a significant reduction (less than $3\sigma$) when combining data from the early and late Universe. We have also observed that the $\beta$ value is sensitive to the combination of data used. On the other hand,  using the DESI-25 data with the CMB and SNIa measurements i.e. PantheonPlus, Union3, or DES-Y5, we obtain a negative value for $\beta$, which means that these combinations favor $w_{\text{de}}>-1$, and the model describes a quintessence dark energy.\\
According to Bayes factors, we find that the LR model provides an improved fit only to CMB data. For all other datasets, we obtain a negative value for $\ln{(\mathcal{B})}$ indicating that our model is disfavoured by CMB combined with BAO, DESI-DR2, or SNIa measurements.

\FloatBarrier

\end{document}